\documentclass[conference]{IEEEtran}

\usepackage[normalem]{ulem}

\usepackage{graphicx}
\usepackage{amsmath}
\usepackage{url}
\usepackage{multirow}
\usepackage{subfig}
\usepackage{authblk}

\begin{document}

\title{10-millisecond Computing}

\author[$\star$]{\normalsize Gang Lu}
\author[$\ast$,$\ddagger$]{Jianfeng Zhan}
\author[$\ast$]{Tianshu Hao}
\author[$\ast$,$\ddagger$]{Lei Wang}

\affil[$\star$]{Beijing Academy of Frontier Science and Technology}
\affil[$\ast$]{Institute of Computing Technology, Chinese Academy of Sciences}
\affil[$\ddagger$]{University of Chinese Academy of Sciences
      \authorcr {lugang@mail.bafst.com, zhanjianfeng@ict.ac.cn, haotianshu@ict.ac.cn, wanglei\_2011@ict.ac.cn}}


\date{}
\maketitle

\begin{abstract}

  Despite computation becomes much complex on data with an unprecedented scale, we argue computers or smart devices should and will \emph{consistently} provide information and knowledge to human being in the order of a few tens milliseconds. We coin a new term 10-millisecond computing to call attention to this class of workloads.

10-millisecond computing raises many challenges for both software and hardware stacks. In this paper, using a a typical workload---memcached on a 40-core server (a main-stream server in near future), we quantitatively measure 10-ms computing's challenges to conventional operating systems. For better communication,  we propose a simple metric---\emph{outlier proportion} to measure quality of service: for $N$ completed requests or jobs, if $M$ jobs or requests' latencies exceed the \emph{outlier threshold } $t$, the outlier proportion is $\frac{M}{N}$.
  For a 1K-scale system running Linux (version 2.6.32), LXC (version 0.7.5 ) or XEN (version 4.0.0), respectively,  we surprisingly find that so as to reduce the service outlier proportion to 10\% (10\% users will feel QoS degradation),   the outlier proportion of a single server has to be reduced by 871X, 2372X, 2372X accordingly.  
  Also, we discuss the possible design spaces of 10-ms computing systems from perspectives of datacenter architectures, networking, OS and scheduling,  and benchmarking.






\end{abstract}

\section{Introduction}



Despite computation becomes much complex on data with an unprecedented scale, in this paper we argue computers or smart devices should and will \emph{consistently} provide information and knowledge to human being in the order of a few tens milliseconds.  We coin a new term 10-millisecond (in short, 10-ms) computing  to call attention to this class of workloads.

First, determined by the nature of human being's nervous and motor systems, the timescale for many human activities is in the order of a few hundreds milliseconds~\cite{dix1996natural,card1983psychology,card1991information}. For example, in a talk, the gaps we leave in speech to tell the other person it is 'your turn' are only a few hundred milliseconds long~\cite{dix1996natural}; the response time of our visual system to a very brief pulse of light and its duration is also in this order.  Second, one of the key results from early work on delays in command line interfaces is that regularity is of the vital importance~\cite{dix1996natural,card1983psychology,card1991information}. If people can predict how long they are likely to wait they are far happier~\cite{dix1996natural,card1983psychology,card1991information}. Third, the experiments in ~\cite{card1983psychology} show perceptual events occurring within a single cycle (of this timescale) are combined into a single percept if they are sufficiently similar, indicating our perceptual system cannot provide much \emph{finer} capability. That is to say, much lower latency (i.e., less than 10 milliseconds) means nothing to human being. So perfect human-computer interactions come from human being's requirements, and should be irrelevant to data scale, task complexity, and their underlying hardware and software systems.

The trend of 10-ms computing has been confirmed by current internet services industries. Internet service providers will not lower their QoS expectation because of the complexity of underlying infrastructures. Actually, keeping latency low is of the vital importance for attracting and retaining users~\cite{dean2013tail, card1991information, Kohavi:2007:PGC}.
Google~\cite{Marissa:2008:keynote} and Amazon~\cite{Kohavi:2007:PGC}  found that moving from a 10-result page loading in 0.4 seconds to a 30-result page loading in 0.9 seconds caused a decrease of 20\% of the traffic and revenue; Moreover delaying the page in increments of 100 milliseconds would
result in substantial and costly drops in revenue.

The trend of 10-ms computing is also witnessed by other ultra-low latency applications~\cite{alizadeh2012less}; for example, high-frequency trading
and internet of thing applications. These applications are characterized by a
request-response loop involving machines in stead of humans,
and operations involving multiple parallel requests/RPCs
to thousands of servers~\cite{alizadeh2012less}. Since a service processing or a  job completes
when all of its requests or tasks are satisfied, the worst-case latency of
the individual requests or tasks is required to be ultra-low  to maintain service or job-level quality of
service. Someone may argue that those applications demand lower and lower latency. However, as there are end-host stacks, NICs (network interface cards),
and switches on the path of an end-to-end application  at which a request or response  currently experience delay~\cite{alizadeh2012less}, we believe in next decade 10-ms is a reasonable latency performance goal for most of end-to-end applications with ultra-low latency requirements.

Previous work~\cite{ousterhout2013case} also demonstrates that it is advantageous to break data-parallel jobs into tiny tasks each of which complete in hundreds of milliseconds. Ousterhout et al.~\cite{ousterhout2013case} demonstrate a 5.2x improvement
in response times due to the use of smaller tasks:
Tiny tasks 
alleviate long wait times seen in today’s clusters for interactive jobs---even large batch jobs can be split into small
tasks that finish quickly.


However, 10-ms computing raises many challenges to both software and hardware stack. In this paper, we quantitatively measure the challenges raised for conventional operating systems. memcached \cite{memcached_site} is a popular in-memory key-value store intended for  speeding up dynamic web applications by alleviating database loads. The average latency is about tens or hundreds $\mu$s. A real-world memcached-based application usually need to invoke several $get$ or $put$ memcached operations, in addition to many other procedures, to serve a single request, so we choose it as a case study on 10-millisecond computing.

Running memcached on a 40-core Linux server,  we found, when the outlier threshold  decreases, the outlier proportion of a single server will significantly deteriorate. Meanwhile, the outlier proportion also deteriorates as the system core number increases. The outlier is further amplified by the system scale. For a 1K-scale system running Linux (version 2.6.32) or LXC (version 0.7.5 ) or XEN (version 4.0.0)---a typical configuration in internet services, we surprisingly find that so as to reduce the service outlier proportion to 10\% (The outlier threshold is 100 $\mu$s), the outlier proportion of a single server needs to be reduced by 871X, 2372X, 2372X, accordingly.  We also conducted a list of experiments to reveal the current Linux systems still suffer from poor performance outlier. The \emph{operating systems} we tested include Linux with different kernels: 1) $2.6.32$, an old kernel released five years ago but still popularly used and in long-term maintenance. 2) $3.17.4$, a latest kernel released on November 21, 2014. 3) $2.6.35M$, a modified version of 2.6.35 integrated with \emph{sloppy counters} proposed by Boyd-Wickizer et al. to solve scalability problem and mitigate kernel contentions \cite{Boyd-Wickizer:2010:MOSBench} \cite{mosbench_site}. 4) representative \emph{real time schedulers}, SCHED\_FIFO (First In First Out) and SCHED\_RR (Round Robin). This observation indicates that the new challenges are significantly different from traditional outlier and stagger issues widely investigated in MapReduce and other environments~\cite{lin2009curse,ibrahim2010leen,kwon2010skew,kwon2012skewtune,ousterhout2013sparrow}.
Furthermore, we discuss the possible design spaces and challenges from perspectives of datacenter architectures, networking, OS and scheduling, and benchmarking.

Section~\ref{PS}  formally states the problem. Section~\ref{OS_perspective} quantitatively measures the OS challenges  in terms of reducing outlier proportion. Section~\ref{design_space} discusses the possible design space of 10-ms computing systems from perspectives of datacenter architectures, networking, OS and Scheduling, and benchmarking. Section~\ref{RW} summarizes the related work. Section~\ref{Conclusion} draws a conclusion.

\section{Problem Statement}~\label{PS}

For scale-out architecture, a feasible solution is to break data-parallel jobs into tiny tasks~\cite{ousterhout2013case}. On the other hand,  for a large-scale online service, a request is often fanned out from a root server to a large number of leaf servers (handling sub-requests) and responses are merged via a request-distribution tree~\cite{dean2013tail}.


We use a probability function $Pr(T\leq t)$ where $T\geq 0$ describes the distribution of service or job-level response time ($T$). If $SC$ $(SC\geq 0)$  leaf servers (or slave nodes) are used to handle sub-requests or tasks sent from the root server (or master node), we use $T_{i}$ to denote the response time of a task or sub-request on server $i$. Here, for clarity, we intentionally ignore the overhead of merging responses from different sub-requests. Meanwhile, for the case of breaking a large job into tiny tasks, we only consider the most simplest scenario----one-round tasks are merged into results, excluding the iterative computation scenarios.

The service or job-level outlier proportion is defined as follows: for $N$ completed requests or jobs, if $M$ jobs or requests' latencies exceed the \emph{outlier threshold} $t$, e.g. 10 milliseconds, the outlier proportion $op\_sj(t)$ is $\frac{M}{N}$.

 According to ~\cite{dean2013tail}, the service or job-level outlier proportion will be extraordinarily magnified by the system scale $SC$.


The outlier proportion of a single server is represented by $op(t)=Pr(T> t)=1-Pr(T\leq t)$.

Assuming the servers are independent from each other, the service or job-level outlier proportion, $op\_sj(t)$, is denoted by Equation~\ref{equation_2}

\begin{align}\label{equation_2}
    op\_sj&(t)=Pr(T_{1}\geq t\ or\ T_{2}\geq t, ...,\ or\ T_{SC}\geq t) \\
    &=1-Pr(T_{1}\leq t)Pr(T_{2}\leq t)...Pr(T_{SC}\leq t) \\
    &=1-Pr(T\leq t)^{SC}=1-(1-Pr(T>t))^{SC}\\
    &=1-(1-op(t))^{SC}
\end{align}

When we deploy the XEN or LXC/Docker solutions, the service or job-level outlier proportion will be further amplified by the number $K$ of guest OSes or containers deployed on each server.

\begin{align}\label{equation_3}
    op\_sj&(t)=Pr(T_{1}\geq t or T_{2}\geq t, ..., or T_{SC*K}\geq t) \\
    &=1-(1-op(t))^{SC*K}
\end{align}






On the vice versa, to reduce an service or job-level outlier proportion to be $op\_sj(t)$, the outlier proportion of a single server must be low as shown in Equation~\ref{equation_4}.

\begin{align}\label{equation_4}
    op(t)=1-\sqrt[SC]{1-op\_sj{(t)}}
\end{align}


For example, a Bing search may access 10,000 index servers~\cite{litales}.  If we need to reduce the service or job-level  outlier proportion to be less than 10\%, the outlier proportion of a single server must be low as $1-(0.9)^{1/10000}\approx0.000011$. Unfortunately, Section~\ref{OS_perspective} shows  it is an impossible task for the conventional OS like Linux to provide such capability.

From a cost-effective perspective, another important performance goal is the valid throughput---how many requests or jobs are finished within the outlier threshold. In fact, according to the outlier proportion, it is very easy to derive the valid throughput. According to the definition of the outlier proportion, $N$ is the throughput number. The valid throughput is $(N-M)$.

Another widely-used metric is \emph{ n\% tail latency}. For the completeness, we also include its definition. The $n\%$ tail latency is the mean latency of all requests beyond a certain percentile $n$~\cite{kasture2014ubik}, e.g., the 99th percentile latency.  Outlier proportion and tail latency are two related concepts, however, there are subtle differences between two concepts.  With respect to the metric $n\%$ tail latency, the outlier proportion is much easier to be used to represent both the user requirement, e.g., $(\frac{N-M}{N})$ requests or jobs satisfying  within the outlier threshold  and the service provider's concern, e.g., the valid throughput $(N-M)$ measuring how many requests
or jobs finished within the outlier threshold. We also note that we cannot derive the valid throughput from the $n$\% tail latency.  As the outlier proportion does not rely upon the history latency data, while the tail latency needs to calculate the average latency of all requests beyond a certain percentile, so the former is easier to handle in the QoS guarantee.  



\subsection{Related concepts}

Soft real time systems are those in which timing requirements are statistically defined~\cite{barabanov1997linux}. An example can be a video conference system---it is desirable that frames are not skipped but it is acceptable if a frame or two is occasionally missed. The goal of a soft real time system  is not to reduce the service or job-level outlier proportion but to reduce the average latency within a specified threshold. If we use the tail latency to describe, that is 0\% tail latency must be less than the threshold. Instead, in a hard real time system, the deadlines must be guaranteed. That is to say, the service or job-level outlier proportion must be zero! For example if during a rocket engine test this engine begins to overheat the shutdown procedure must be completed in time~\cite{barabanov1997linux}.

Different from hard real time systems, 10-ms computing systems not only care about performance outlier---specifically outlier proportion, but also the resource utilization, while the former often uses dedicated systems to achieve the zero outlier proportion goal without paying much attention to the resource utilization.

\begin{figure}[t]
  \centering
  \label{outlier_proportion_limit}
  \includegraphics[width=0.48\linewidth]
  {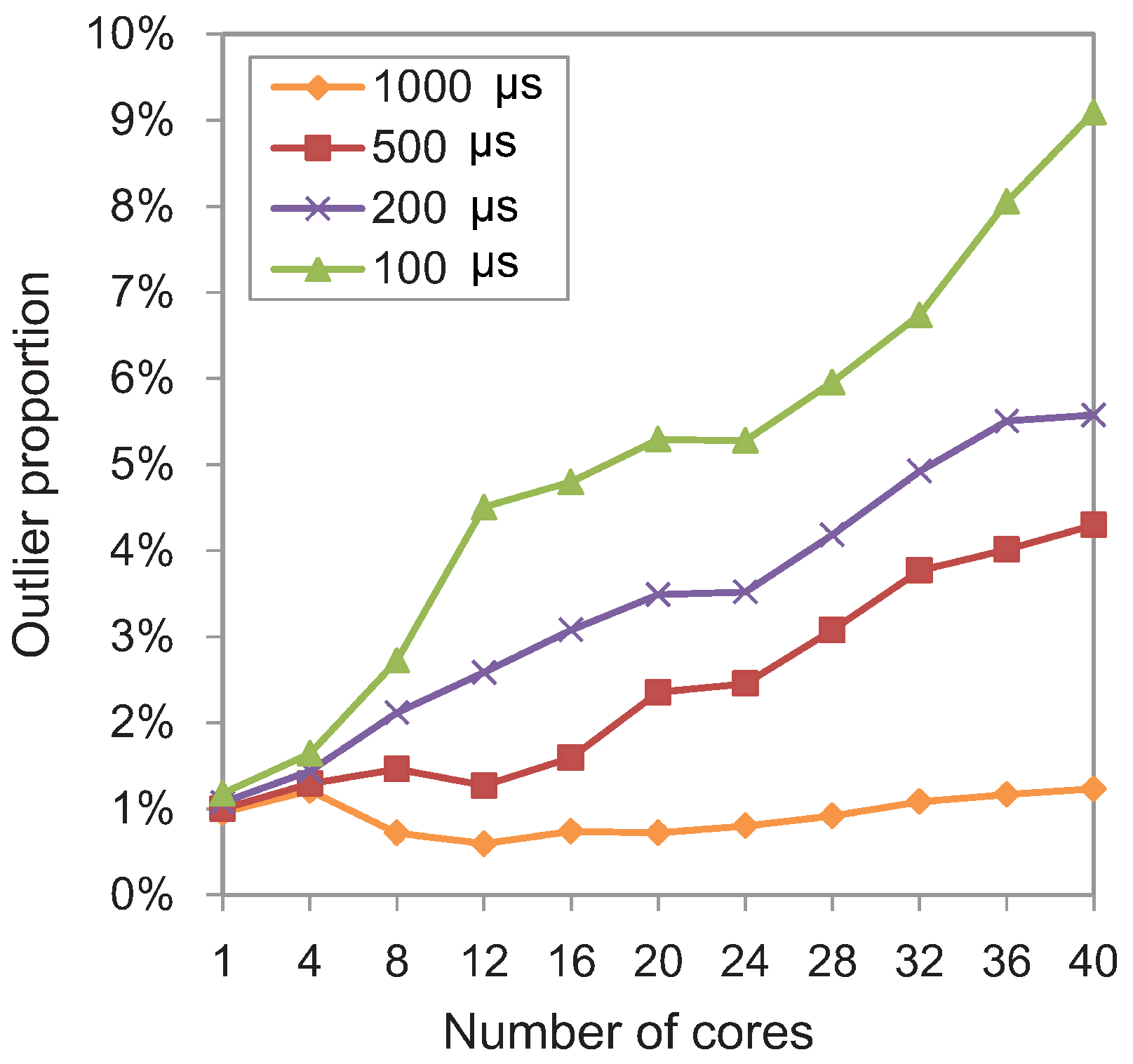}
  \caption{The outlier proportion of memcached on each server changes with the outlier thresholds (from 100 to 1000 $\mu$s) and core numbers (from 1 to 40).}
  \label{outlier_proportion_limit}
\end{figure}

\section{Challenges from an OS perspective}~\label{OS_perspective}
We investigate the outlier problem from a perspectives of the operating system using a latency-critical workload: memcached. memcached \cite{memcached_site} is a popular in-memory key-value store intended for  speeding up dynamic web applications by alleviating database loads. The average latency is about tens or hundreds $\mu$s. A real-world memcached-based application usually need to invoke several $get$ or $put$ memcached operations, in addition to many other procedures, to serve a single request, so we choose it as a case study on 10-millisecond computing. We increase the running cores and bind a memcached instance on each active core via \emph{numactl}. A 40-core ccNUMA (cache coherent Non Uniform Memory Access) machine with four NUMA domains (Each with a 10-core E7-8870 processor.) is used for running memcached instances. Four 12-core servers run client emulators, which are obtained from MOSBench \cite{Boyd-Wickizer:2010:MOSBench}. The host OS is SUSE11SP1 with the kernel version 2.6.32 and a default scheduler CFS (Completely Fair Scheduler).

Figure~\ref{outlier_proportion_limit} shows the outlier proportions vs. different outlier thresholds and increasing cores. We can observe that: a) tighter outlier thresholds lead to higher outlier proportions. The  outlier proportion is low as of 0.60\% on a common 12-core node with the outlier threshold of 1 second. By contrast, it will be high as 4.50\% if we reduce the outlier threshold to 100-$\mu$s. b) The outlier proportion increases gradually with the number of cores. In the worst cases, it degrades to to 9.09\%. According to Equation~\ref{equation_2}, using 1K 40-core servers, the service or job-level outlier proportion will be up to nearly 100\%.

Following the above observations, we quantitatively measure  the challenges in terms of reducing outlier proportion using a monolithic kernel (Linux) and virtualization technologies including Xen and Linux Containers.

\begin{figure}[t]
  \centering
  \subfloat[]{\label{Linux_outlier_proportion_comparison}   \includegraphics[width=0.48\linewidth]    {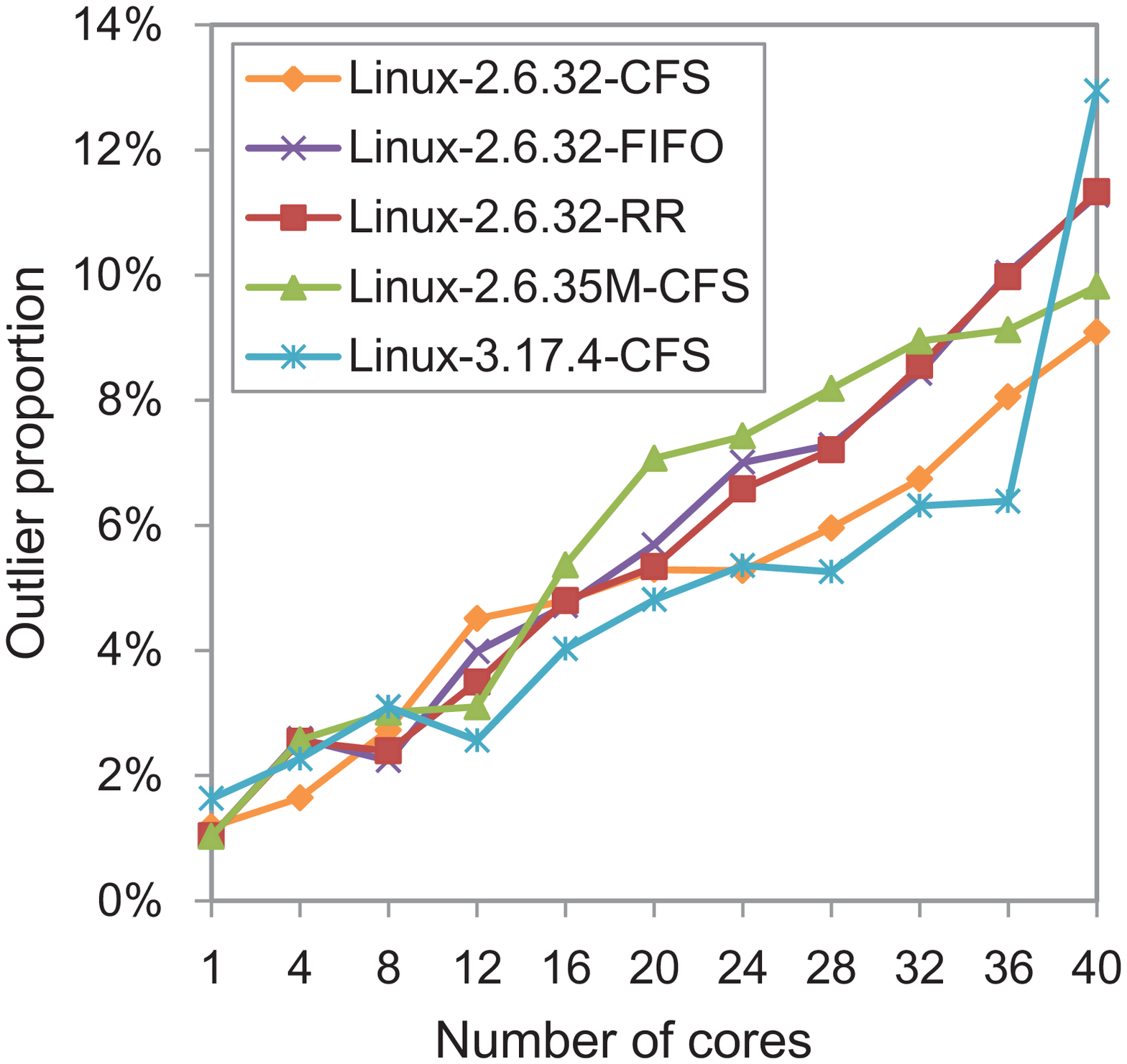}}
  \subfloat[]{\label{Linux_outlier_throughput_comparison}   \includegraphics[width=0.48\linewidth]
  {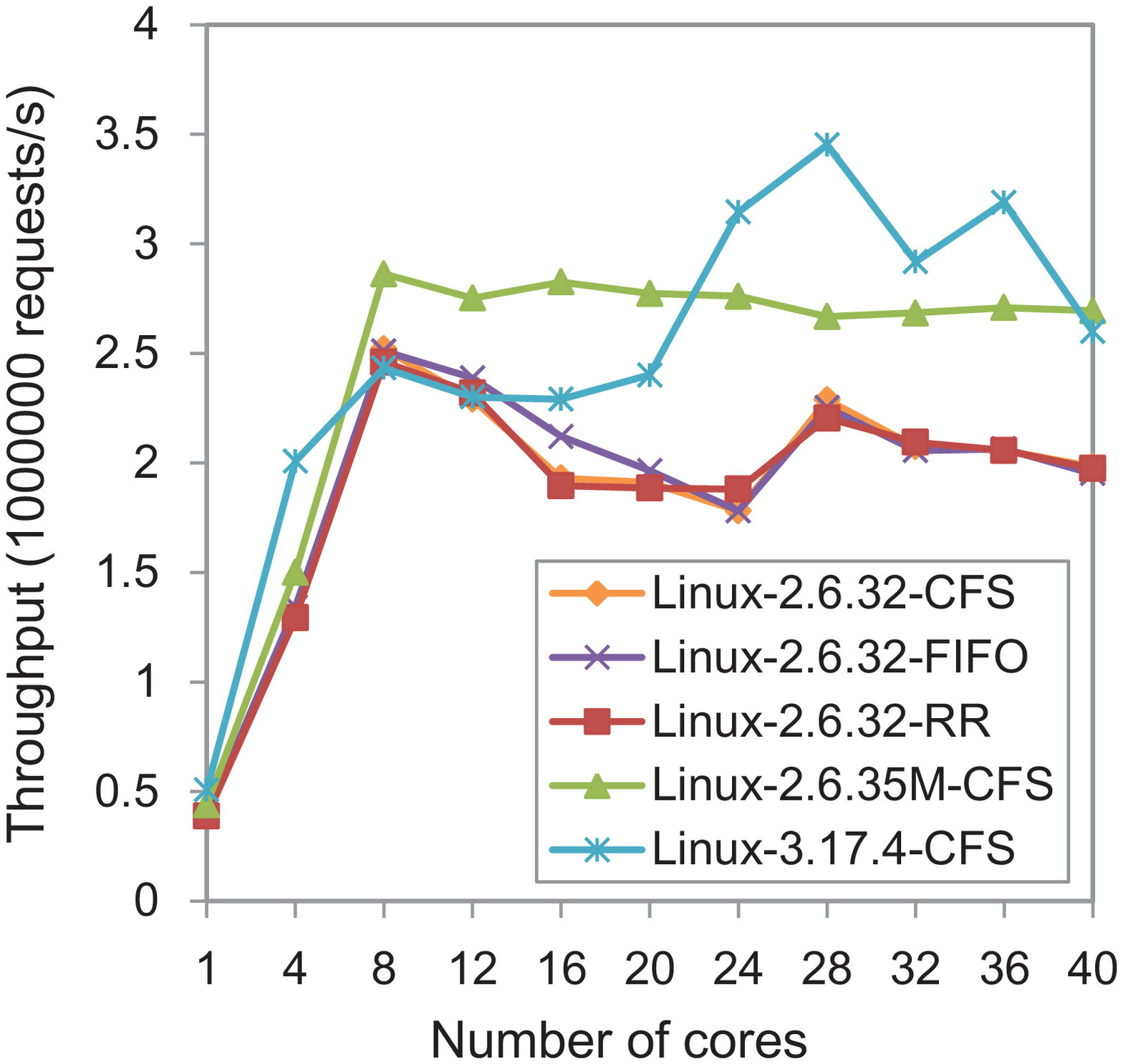}}
  \caption{(a) shows the outlier proportions of a single server using different Linux kernels and schedulers. (b) shows the valid throughput. The core number of a server is varied (see X axis). The outlier threshold is 100 $\mu$s.}
  \label{challenge_outlier_proportion_linux}
\end{figure}
\subsection{A monolithic kernel: Linux}
We conducted a list of experiments to study whether current Linux systems still suffer from poor outlier performance. The \emph{operating systems} we tested are Linux with different kernels: 1) $2.6.32$, an old kernel released five years ago but still popularly used and in long-term maintenance. 2) $3.17.4$, a latest kernel released on November 21, 2014. 3) $2.6.35M$, a modified version of 2.6.35 integrated with \emph{sloppy counters} proposed by Boyd-Wickizer et al. to solve scalability problem and mitigate kernel contentions \cite{Boyd-Wickizer:2010:MOSBench} \cite{mosbench_site}. \emph{sloppy counters} adopts local counters on each core and an central counter to avoid unnecessary touching of the global reference count. Their evaluations show its good effects on mitigating unnecessary contentions on kernel objects. Beside these systems with different kernels, we also evaluated the impact of representative \emph{real time schedulers}, SCHED\_FIFO (First In First Out) and SCHED\_RR (Round Robin). A SCHED\_FIFO process runs until either it is blocked by an I/O request (if a higher priority process is ready) or itself invokes \emph{sched\_yield}. SCHED\_RR is a simple implementation of SCHED\_FIFO, except that each is allowed to run for a maximum time slice. The time slice in SCHED\_RR is set to 100 $\mu$s. We use a dash with CFS, FIFO, or RR following the kernel version to denote the scheduler the kernel uses in Figures.

\emph{Results and discussions}. Figure~\ref{Linux_outlier_proportion_comparison} and Figure~\ref{Linux_outlier_throughput_comparison} show that these systems are not competent for low outlier proportion under the outlier threshold of 100 $\mu$s. First, all systems has a scalability problem in terms of the valid throughput. Although the modified kernel Linux-2.6.35M behaves better than Linux-2.6.32, the throughput stops increasing after 8 cores. Second, the outlier proportions climb up to ~10\% with 40 cores. Besides, the latest kernel 3.17.4 still cannot surpass the older kernel 2.6.35M. Such a ad-hoc method of mitigating potential resource contentions seems to be endless and contribute to limited improvements. Third, the real time schedulers neither reduce the outlier proportion nor improve the valid throughput. The real time schedulers do not show positive effects on the performance.


\begin{figure}[t]
  \centering

  \subfloat[ ]{\label{challenge_outiler_proportion_lxc_xen}   \includegraphics[width=0.48\linewidth]
  {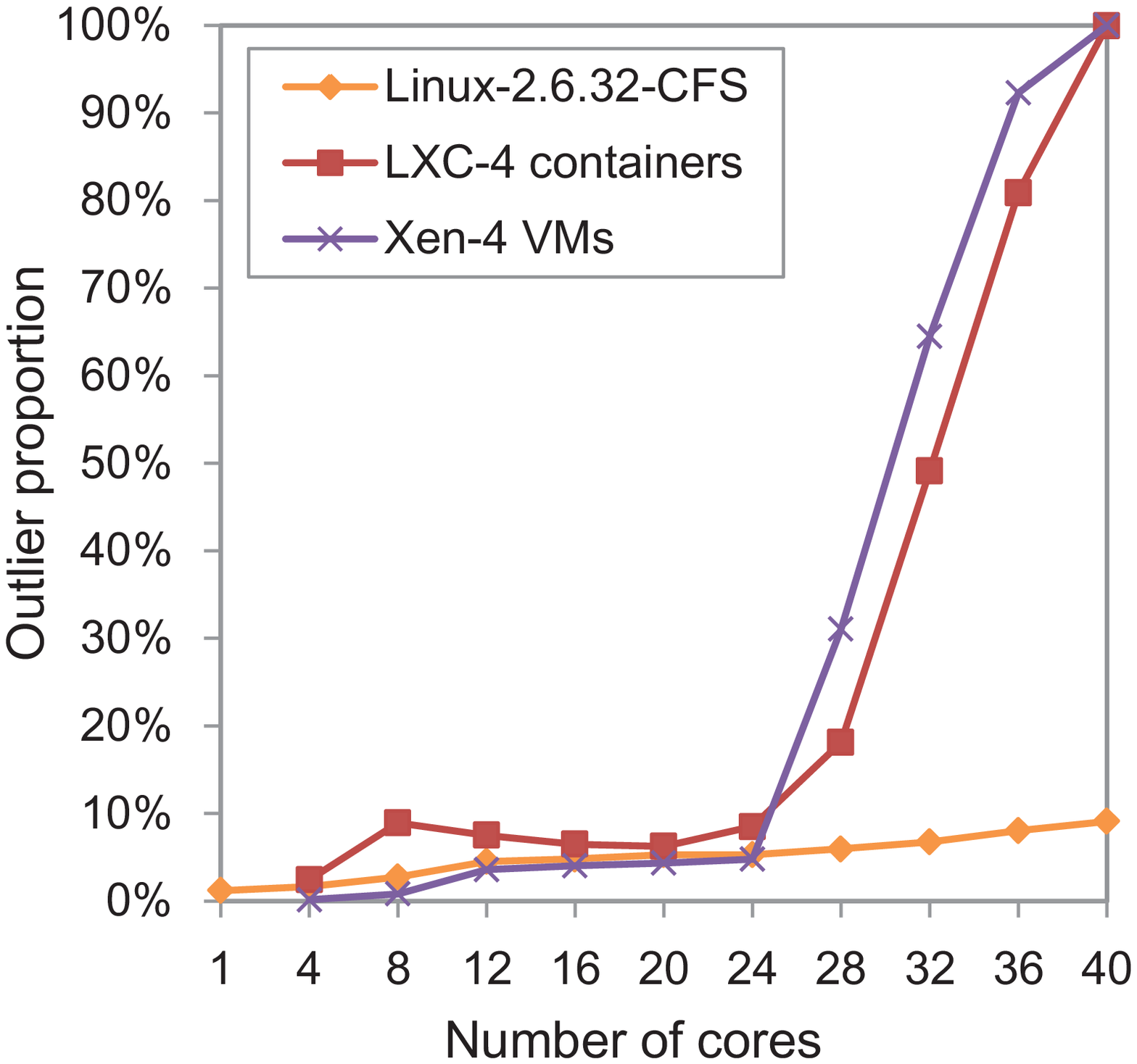}}
  \subfloat[]{\label{challenge_1000node}   \includegraphics[width=0.48\linewidth]
  {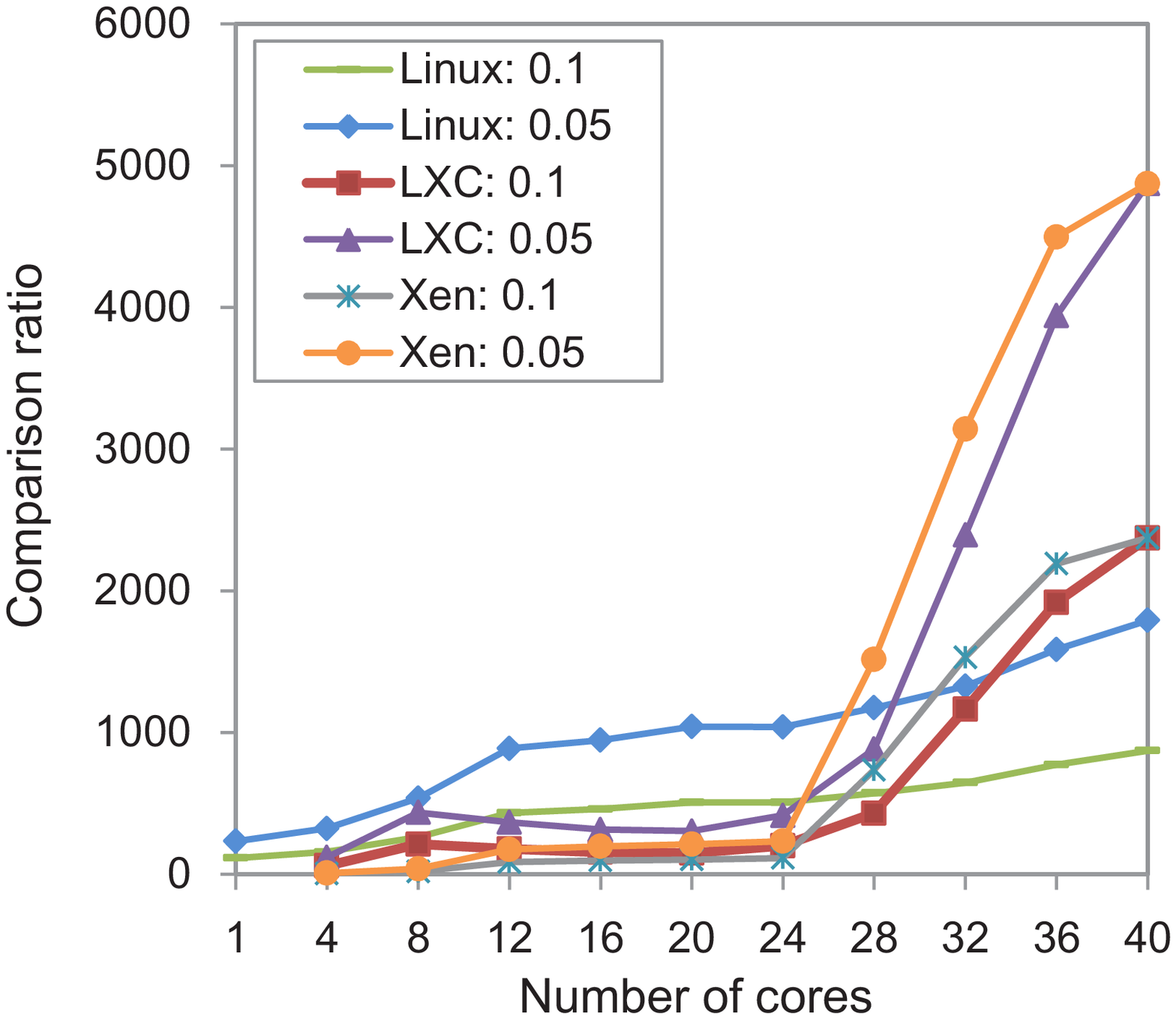}}
  \caption{{(a) shows the outlier proportions of a single server when running one Linux instance, 4 containers (LXC) and 4 VMs (XEN) on the varied core number (see X axis). The outlier threshold is 100 $\mu$s.  (b) shows for 1K-server configuration, how many times of the outlier proportion (see Y axis) of each server  need to be reduced to reduce the service or job-level outlier proportion to 10\% or 5\%. The core number of a server is varied (see X axis). The outlier threshold is 100 $\mu$s.}}
  \label{challenge_lxc_xen}
\end{figure}

\subsection{Virtualization technologies}
Virtualization offers multiple execution environments on a single server. Comparing with the monolithic kernel OS,
according to Equation~\ref{equation_3}, the number of guest OS or containers will amplify the outlier, and hence make the outlier proportion deteriorates.
We use LXC~\cite{soltesz2007container} and Xen\cite{chisnall2008definitive}  to evaluate the outlier
proportions.  The versions of LXC and Xen are 0.7.5 and 4.0.0, respectively. The VMs on Xen are all para-virtualized.
For both LXC and Xen, the node resources are divided equally for the 4 containers and 4 VMs.


LXC is based on a monolithic kernel which binds  multiple processes (process group) together to run as a full-functioned OS. A container-based system can spawn several shared, virtualized OS images, each of which is known as a container. It consists of multiple processes bound together (process group), appearing as a full-functioned OS with an exclusive root file system and a safely shared set of system libraries. For Xen-like hardware level virtualization, there is a microkernel-like hypervisor to manage all virtual machines (VMs). Each VM is composed by virtual devices generated by device emulators and runs on a less privileged level. The execution of privileged instructions on a VM must be delivered to hypervisor. Communications are based on the event channel mechanism.



We run memcached instances  on four containers and four virtual machines (VMs) hosted on a single node, respectively. A request is fanned out to the four containers or VMS and four sub-query responses are merged in the client emulator to calculate the performance. From Figure~\ref{challenge_outiler_proportion_lxc_xen}, we observe that the performance is far from the expectation.  Note that when deploying on less than 24 cores, Xen has better outlier proportions. Comparing to LXC, Xen has much better performance isolation.  Xen's and LXC 's outlier proportion significantly deteriorate, respectively  when each VM and container runs on 10 cores.  We also note that the valid throughput of Xen is much lower than LXC and Linux.

\subsection{Discussion}

 From Figure~\ref{challenge_1000node}, we surprisingly find that to reduce the service or job-level outlier proportion to 10\%, running Linux or LXC or XEN, respectively, the outlier proportion of a single server needs to be reduced by 871X, 2372X, 2372X accordingly. And if reducing the outlier proportion to 5\%, the performance gap is 1790X, 4721X, 4721X, respectively.

The operating system can be abstracted as a multi-thread scheduling system
with internal and external interrupts. The outliers are mainly caused by the tasks
starved for certain resources because of preemption or interferences from
parallel tasks or kernel routines. Waiting and serialization can be
aggravated by the increasing parallel tasks and cores. Here are a few
occasions that the outlier proportion may be aggravated with increasing cores.

\begin{itemize}
\item \textbf{Synchronization}. Synchronization becomes more frequent and
time-consuming, such as the lock synchronization of resource
counters, cache coherence and TLB shootdown among cores. For example,
for a multiprocessor OS, TLB consistency is maintained through by
invalidating entries when pages are unmapped. Although the TLB
invalidation itself is fast, the process of context switches,
transferring IPIs (Inter-Processor Interrupts) across all possible
cores, and waiting for all acknowledgements may be time-consuming
\cite{Villavieja:2011:DiDi}. On one hand, processing shootdown IPIs
need to interrupt current running tasks. On the other hand, the time
consumed during transferring and waiting may easily climb rapidly with
the increasing cores.

\item \textbf{Shared status}. Shared status also becomes more common, such as shared queues and lists or shared buffers. For example,
information of software resources or kernel states are commonly stored in
queues, lists, or trees, such as the per-block list that tracks open
files. With increasing tasks on more cores, these structures may be more
filled. Thus, searching and traversing them becomes more expensive.
Besides, there are many limits set by the kernel. In 10-ms computing, the
number of tiny tasks may be of large quantity whose accesses of system
resources may be easily excessive.

\item \textbf{Scheduling based on limited hardware}. Hardware resources
such as last level cache (LLC), inter-core interconnect, DRAM, I/O hub
are physically shared by all processors. Too many operations on the
subsystems may reach beyond the capacity, so it is difficult for a
scheduler to schedule tasks with a optimum solution.

\end{itemize}

\section{Possible Design Space and Challenges}~\label{design_space}

This section discusses possible design space and challenge from perspectives of datacenter architecture, networking, OS and Scheduling, and benchmarking.

\subsection{Data center architecture}

Existing datacenters are built using commodity servers. A common 1000-node cluster often come with a significant performance penalty because
networks at datacenter scales have historically been oversubscribed---communication
was fast within a rack, slow off-rack, and worse still for
nodes whose nearest ancestor was the root~\cite{nightingale2012flat}.
The datacenter bandwidth shortage forced software developers to think in terms of ¡°rack locality¡±---
moving computation to data rather than vice versa~\cite{nightingale2012flat}.  The same holds true for the storage systems.
In this server-centric setting, resource shortage will result in excessive competition and  significantly deteriorate performance outliers for 10-ms computing.

Recent efforts suggest a forthcoming paradigm
shift towards a disaggregated datacenter (DDC), where
each resource type is built as a standalone resource pool
and a network fabric interconnects these resource pools~\cite{gao2016network}.

On one hand, this paradigm shift is driven
largely by hardware architects.
CPU, memory and
storage technologies exhibit significantly different trends in
terms of cost, performance and power scaling, and hence it is increasingly hard to adopt evolving
resource technologies within a server-centric architecture~\cite{gao2016network}.
By decoupling CPU, memory and storage
resources, DDC makes it easier for each resource technology
to evolve independently and reduces the time-to-adoption
by avoiding the burdensome process of redoing integration
and motherboard design~\cite{gao2016network}.

On the other hand, as the resource is much uniformly accessed with respect to the traditional server-centric architecture, DDC helps alleviate performance outliers. In addition, it
also enables fine-grained and efficient individual resource provisioning and
scheduling across jobs~\cite{han2013network}.

Unfortunately, resource disaggregation will further challenge networking, since disaggregating CPU from memory
and disk requires that the inter-resource communication
that used to be within the scope of  a server must now traverse
the network fabric. Thus, to support good application-level
performance it becomes critical that the network fabric
provide much lower latency communication~\cite{han2013network}.


\subsection{Networking}

In an end-to-end application within a  datacentre, there are end-host stacks, NICs (network interface cards),
and switches at which packets experience delay~\cite{alizadeh2012less}.  Kernel bypass and zero
copy techniques~\cite{cohen2009remote} ~\cite{shivam2001emp}  significantly reduce the latency at
the end-host and in the NICs.

Rumble et al~\cite{rumble2011s} argue that it should be possible
to achieve end-to-end remote procedure call (RPC)
latencies of 5-10$\mu$s in large datacenters using commodity
hardware and software within a few years. However,
achieving that goal requires creating a
new software architecture for networking with a different
division of responsibility between operating system,
hardware, and application.  Over the longer-term, they also think 1$\mu$s datacenter round-trips can be
achieved  with more radical
hardware changes, i.e., moving the NIC onto the
CPU chip.


Trivedi et al~\cite{trivedi2016ir} also found that the current Spark data processing
frameworks can not leverage network speeds higher than 10
Gbps because the high amount of CPU work done
per byte moved eclipses any potential gains coming from
the network.   This observation indicates 10-ms computing need more than only ultra-low latency networking, and we need investigate a number of ways to balance
CPU and network  time.




\subsection{OS and scheduling}

The widely
used Linux, Linux container (LXC)~\cite{banga1999resource}, or Xen~\cite{barham2003xen}, adopts
a monolithic kernel or a virtual machine monitor (VMM)
that shares numerous data structures protected by locks (share
first), and then a process or virtual machine (VM) is proposed
to guarantee performance isolation (then isolate).  These  OS architectures have their
 inherent structure obstacle in reducing outlier proportion,
because globally-shared and contended data structures and maintaining global coordination deteriorate performance
interferences and outliers, especially for 10-ms computing. Our evaluation in Section~\ref{OS_perspective}
shows running with three Linux kernels---2.6.32, 2.6.35, 3.17.4, respectively, on a 40-core server, memcached exhibits deteriorated outlier proportion as the core number increases, while its average
latency does not increase much.

The shift toward 10-ms computing calls
for new OS architectures for reducing the outlier proportion.
Our previous work~\cite{lu2016horizontal} presents an "isolate first, then share" (in short,
IFTS) OS model that is promising in guaranteeing worst-case performance. We decompose the
OS into the supervisor and several subOSes running in parallel.
A subOS isolates resources first through directly managing
physical resources without intervention from the supervisor,
and the supervisor enables resource sharing through
creating, destroying, resizing a subOS on-the-fly; SubOSes
and the supervisor have confined state sharing, but fast inter-subOS communication mechanisms based on shared memory
and IPIs are provided on demand.
On several Intel Xeon platforms, we applied the IFTS OS
model to build the first working prototype---RainForest. We
ported Linux 2.6.32 as a subOS and performed comprehensive
evaluation of RainForest against three Linux kernels: 1)
2.6.32; 2) 3.17.4; 3) 2.6.35M; LXC (version 0.7.5 );
Xen (version 4.0.0).
Experimental results show: 1)With respect to Linux, LXC,
and Xen, RainForest improves the throughput of the search
service by 25.0\%, 42.8\%, 42.8\% under the 99th percentile latency
of 200 ms.
The CPU utilization rate is improved by 16.6\% to
25.1\% accordingly.
Our previous work shows achieving better isolation among OS instances
will provide better worst-case performance and is promising for 10-ms computing workloads.

In addition to OS support, supporting 10-ms computing requires a low-latency, high
throughput task scheduler. As Ousterhout et al. demonstrate in~\cite{ousterhout2013case}, handling 100ms tasks in a
cluster with 160.000 cores (e.g., 10,000 16-core machines),
requires a scheduler that can, on average, make
1.6 million scheduling decisions per second. Today¡¯s
centralized schedulers have well-known scalability limits~\cite{schwarzkopf2013omega}
 that hinder their ability to support millisecond-scale tasks in a
large cluster. Instead, handling large clusters and very short tasks
will require a decentralized scheduler design like Sparrow~\cite{ousterhout2013sparrow}.
In addition to providing high throughput scheduling
decisions, a framework for 10-ms computing must also reduce
the overhead for launching individual tasks~\cite{ousterhout2013case}.

\subsection{Benchmarking}
To perform research on 10-ms computing, the first challenge is to set up a benchmark suite. Unfortunately, it will be a non-trivial issue as 10-ms computing may depend on the new scheduling, executing and programming models. So the challenge lies in  how to  set up a benchmark suite on the  existing commodity software system as we have not implemented the new scheduling, executing and programming models for 10-ms computing. Previous work like BigDataBench~\cite{Wang:2014:BigDataBench} or TailBench ~\cite{kasture2016tailbench} still has serious drawback. For example, TailBench~\cite{kasture2016tailbench} only provides several simple workloads that have latencies varying from microseconds to seconds.

\section{Related Work}~\label{RW}

The outlier problem has been studied in many areas such as parallel iterative convergent algorithms where all executing threads must be synchronized
\cite{cipar2013solving}. Within the context of scale-out architecture, we now discuss related work on outlier sources and mitigation.



\subsection{Sources of Outliers and Tail Latency}


\emph{Hadoop MapReduce Outliers}. The problem of Hadoop outliers is first proposed in \cite{dean2008mapreduce} and it is further studied in heterogeneous environments \cite{zaharia2008improving}. In Hadoop, the task outliers are typically incurred by task skews, including load imbalance among servers, uneven data distribution, and unexpected long processing time \cite{lin2009curse,ibrahim2010leen,kwon2010skew,kwon2012skewtune,ousterhout2013sparrow}

\emph{Tail latency in interactive services}. In today's WCSs, interactive services and batch jobs are typically co-located on the same machine to increase machine utilizations. In such shared environments, resource contention and performance interference is a major source of service time variability \cite{shue2012performance,flajslik2013network}. This variability is further significantly amplified when considering requests' queueing delay, thus incurring  high request tail latency.


\subsection{Application-level techniques to mitigate outliers}

\textbf{Task/sub-request redundancy} is a commonly applied technique to mitigate outliers and tail latency. The key idea of such technique is to execute each individual task/sub-request in multiple replicas so as to reduce its latency by using the quickest replica. In \cite{ananthanarayanan2013effective,vulimiri2012more}, this technique has been applied to mitigate outlier tasks in small Hadoop jobs whose number of tasks is smaller than ten. In \cite{stewart2013zoolander}, it is applied to reduce low tail latency only when the system at idle state.
In contrast, \textbf{task/sub-request reissue} is a conservative redundancy technique \cite{ananthanarayanan2010reining,jalaparti2013speeding,ousterhout2013sparrow}. This technique first sends a task/sub-request replica to one approximate machine. The replica is judged as the outlier if it is not completed after a short delay (e.g. the 99th percentile latency for this class of tasks/sub-requests), and then a secondary replica is sent to another machine. Both techniques work well when the service is under light load. However, when load becomes heavier, the unnecessarily execution of the same tasks/sub-requests adversely increases the outlier proportion \cite{shah2013redundant}.

\textbf{Partial execution} is another widely used technique to mitigate outliers by sacrificing result quality (e.g. prediction accuracy in classification or recommendation services). Following the anytime framework initially proposed in AI \cite{zilberstein1996using}, this technique has been applied in Bing search engine \cite{jalaparti2013speeding,he2012zeta} to return an intermediate and partial search result whenever the allocated processing time expires. Similar approaches have been proposed to sample a subset of input data to produce approximate results for MapReduce jobs under both time and resource constraints \cite{laptev2012early,pansare2011online,shi2012you,wang2014sampling}.
Moreover, best-effort scheduling algorithms have been developed to form a compliment to the partial execution technique, which allocate available processing times among tasks/sub-requests to maximize their result quality \cite{he2011tians,he2012zeta}. However, when load become heavier, such technique incurs considerable loss in result quality to meet response target and this is sometimes unacceptable for users.

\section{Conclusion}~\label{Conclusion}

In this paper we argue computers or smart devices should and will \emph{consistently} provide information and knowledge to human being in the order of a few tens milliseconds despite computation becomes much complex on data with an unprecedented scale. We coin a new term 10-millisecond computing to call attention to this class of workloads.


 We specifically investigate 10-ms computing's challenges raised for conventional operating systems.
 For a 1K-scale system---a typical internet service configuration---running Linux (version 2.6.32) or LXC (version 0.7.5 ) or XEN (version 4.0.0), respectively, we surprisingly find that to reduce the service-level outlier proportion of a typical workload---memcached to 10\%,   the outlier proportion of a single server needs to be reduced by 871X, 2372X, 2372X accordingly.   We also conducted a list of experiments to reveal the state-of-the-art and state-of-the-practice Linux systems still suffer from poor performance outlier, including Linux kernel versions $3.17.4$,  $2.6.35M$, a modified version of 2.6.35 that is integrated with \emph{sloppy counters}, and two representative real time schedulers. This observation indicates the new challenges are significantly different from traditional outlier and stagger issues widely investigated in MapReduce and other environments.  Also, we discuss the possible design spaces and challenges for 10-ms computing systems from perspectives of datacenter architecture, networking, OS and scheduling, and benchmarking.

\bibliographystyle{plain}

\end{document}